\documentclass[conference]{IEEEtran}
\IEEEoverridecommandlockouts
\usepackage{cite}
\usepackage{amsmath,amssymb,amsfonts}
\usepackage{algorithmic}
\usepackage{graphicx}
\usepackage{textcomp}
\usepackage{tabularx}
\usepackage{ltxtable}
\usepackage[table]{xcolor}  
\usepackage{balance}
\usepackage{caption}
\usepackage{subcaption}

\def\BibTeX{{\rm B\kern-.05em{\sc i\kern-.025em b}\kern-.08em
    T\kern-.1667em\lower.7ex\hbox{E}\kern-.125emX}}

\newcommand{\zv}{{\boldsymbol{z}}}
\newcommand{\cv}{{\boldsymbol{c}}}
\newcommand\pname[1]{SAIL}

\begin{document}

\title{SAIL: Unsupervised Spatial-Angular Interpretable Feature Learning for RF Map Synthesis}

\author{\IEEEauthorblockN{Sopan Sarkar and Marwan Krunz}
\IEEEauthorblockA{Department of Electrical and Computers Engineering, University of Arizona, Tucson, USA \\
\{sopansarkar, krunz\}@arizona.edu}
}

\maketitle

\begin{abstract}
In wireless networks, radio-frequency (RF) maps are critical for tasks such as capacity planning, coverage estimation, and localization. 
Traditional approaches for obtaining RF maps, including site surveys and ray-tracing simulations, are labor-intensive or computationally expensive, especially at high frequencies and dense network deployments.
Generative AI offers a promising alternative for RF map synthesis.
However, supervised methods are often infeasible due to the lack of reliable labeled training data, while purely unsupervised methods typically lack explicit control over the synthesis process.
To address these challenges, we propose \textbf{SAIL} (\textbf{S}patial--\textbf{A}ngular \textbf{I}nterpretable \textbf{F}eature \textbf{L}earning), a generative adversarial network (GAN)-based framework that learns interpretable and controllable latent variables directly from unlabeled RF maps and enables targeted RF map synthesis at inference time through latent-variable manipulation.
SAIL builds on the information-maximizing GAN (InfoGAN) principle to learn a structured representation comprising: (i) a categorical latent variable that captures discrete floor-plan regions associated with Tx location and (ii) a continuous latent variable that captures angular variations corresponding to the Tx boresight angle, without requiring any location or orientation supervision during training.
We further adopt a Wasserstein GAN objective with a gradient penalty to improve training stability and synthesis quality. 
Our results using ray-tracing-based RF maps indicate that SAIL learns physically meaningful spatial--angular factors and enables fast controlled RF map synthesis, achieving an average SSIM of 0.8576 and an average PSNR of 23.33~dB relative to ray-tracing simulations.
\end{abstract}


\section{Introduction}

Radio-frequency (RF) maps represent a spatial field of location-dependent radio metrics, such as received signal strength (RSS), path loss, signal-to-noise ratio (SNR) etc., over a physical environment.
They play a critical role in radio network planning, supporting tasks such as coverage estimation, capacity analysis, access point placement, and localization~\cite{athanasiadou2020radio}. 
Accurate RF maps are therefore essential for ensuring quality-of-service guarantees in both cellular and managed Wi-Fi networks.
In practice, RF maps are commonly obtained through site surveys, where engineers collect RSS or SNR measurements at multiple locations within the coverage area. 
Such surveys are labor-intensive and time-consuming, as they require manual measurements while accounting for buildings, walls, furniture, and interference sources.
The challenge is further exacerbated at millimeter-wave (mmWave) frequencies, where coverage remains strongly dependent on Tx location and beam orientation~\cite{rangan2014millimeter}, resulting in highly variable coverage patterns that are difficult to characterize exhaustively.

Ray-tracing simulators provide a physics-based alternative to site surveys for generating site-specific RF maps by numerically approximating electromagnetic propagation in a given environment using geometric descriptions (e.g., floor plans or 3D models) and material electromagnetic parameters.
When the environment and material models are sufficiently accurate, ray-tracing can produce high-fidelity simulated RF maps~\cite{Lecci2021}.
However, ray-tracing complexity grows rapidly with the number of candidate Tx locations, antenna configurations/orientations, and interaction mechanisms (e.g., reflections, diffractions, and transmissions), making exhaustive exploration of large configuration spaces computationally prohibitive~\cite{romero2022}.
Moreover, neither site surveys nor ray-tracing simulations can feasibly enumerate all possible deployment scenarios, leading to incomplete RF map datasets that limit downstream analysis.

These challenges have motivated the development of data-driven RF map synthesis techniques aimed at augmenting measured or simulated datasets.
More recently, deep generative models, including generative adversarial networks (GANs) and diffusion-based models, have emerged as powerful frameworks for synthesizing realistic data samples.
These techniques have been applied to a variety of wireless communication problems, including anomaly detection~\cite{zhou2021radio}, modulation classification~\cite{TangAccess2018}, and channel estimation~\cite{balevi2021wideband,diff_channel_estimate}.
Several studies have also investigated generative models for RF map generation and augmentation, often using conditional architectures that rely on side information such as Tx location or partial measurements~\cite{Seong2020,liu2020cgan,njima2021indoor,rm-gen}.
Despite their promise, existing generative approaches exhibit two key limitations.
First, supervised and conditional models depend on accurate labels for Tx configuration parameters such as Tx location or beam boresight angle, which are often unavailable, noisy, or incomplete in practice.
Second, fully unsupervised models typically lack explicit mechanisms for controlling the synthesis process, making it difficult to generate RF maps corresponding to specific Tx configurations.
As a result, no existing approach simultaneously achieves label-free training and physically meaningful control over key transmitter configuration parameters, limiting the applicability of current methods to network planning and systematic exploration of deployment configurations.

In this paper, we propose SAIL (Spatial--Angular Interpretable Feature Learning), a GAN-based framework that addresses both challenges by discovering and aligning interpretable latent variables with physically meaningful propagation factors.
SAIL builds on the information-maximizing GAN (InfoGAN) formulation~\cite{Chen2016a} to learn an interpretable latent representation of RF maps in a fully unsupervised manner.
Specifically, SAIL learns latent factors that capture (i) spatial variations associated with Tx location and (ii) angular variations associated with the Tx boresight direction.
Although these factors are learned without labeled supervision during training phase, they enable controlled RF map synthesis at inference time through direct latent-variable manipulation.
Our main contributions are as follows:
\begin{itemize}
\item \textbf{Unsupervised spatial--angular disentanglement:}
We introduce a generative framework that discovers and disentangles Tx location and Tx boresight direction directly from unlabeled RF maps, without requiring any location or orientation annotations.

\item \textbf{Controllable RF map synthesis:}
We show that the learned latent variables correspond to physically meaningful spatial and angular factors and act as explicit control parameters at inference time, enabling targeted RF map generation for specific Tx configurations.

\item \textbf{Stable InfoGAN-based training for RF maps:}
We integrate a Wasserstein GAN objective with gradient penalty (WGAN-GP) into SAIL to improve training stability and synthesis fidelity for RF map generation.

\item \textbf{Validation using ray-tracing RF maps:}
We validate SAIL using ray-tracing--simulated RF maps and demonstrate high-fidelity synthesis, achieving an average Structural Similarity Index Measure (SSIM) of 0.8576 and an average Peak Signal-to-Noise Ratio (PSNR) of 23.33~dB relative to ray-tracing simulations.
\end{itemize}

The remainder of this paper is organized as follows. Section~\ref{sec:Preliminaries} reviews the InfoGAN framework.
Section~\ref{sec:SAIL} presents the proposed SAIL architecture and training procedure.
Section~\ref{sec:dataset} describes the RF map dataset generation.  
Section~\ref{sec:Evaluation} reports performance evaluation results, followed by concluding remarks in Section~\ref{sec:conclusions}.
\section{Preliminaries: Introduction to InfoGAN}
\label{sec:Preliminaries}

A basic GAN~\cite{goodfellow2014} consists of two neural networks trained in an adversarial manner: a generator $G$ and a discriminator $D$. 
$G$ maps a random noise vector $\zv \in \mathbb{R}^k$, drawn from a prior distribution $P_{\zv}$ (typically a standard Gaussian or uniform distribution), to a synthetic data sample $G(\zv)$.
$D$ takes a data sample as input and outputs a scalar $D(x)\in[0,1]$ representing the probability that $x$ originates from the true data distribution rather than being generated by $G$.
Let $x$ denote a real data sample drawn from the true data distribution $P_{\text{data}}$, and let $P_g$ denote the distribution induced by $G$.
The GAN training objective is formulated as the following minimax game:
\begin{multline}
\min_{G}\max_{D}\; V(G,D)
\overset{\scriptstyle{\text{def}}}{=}
\mathbb{E}_{x\sim P_{\text{data}}}\big[\log D(x)\big] \\
+ \mathbb{E}_{\zv\sim P_{\zv}}\big[\log(1-D(G(\zv)))\big],
\label{eq:minimaxlogGAN_prelim}
\end{multline}
where $G$ aims to generate samples that are indistinguishable from real data, while $D$ attempts to correctly classify real and generated samples.
At equilibrium, the generator distribution matches the data distribution ($P_g = P_{\text{data}}$), and the discriminator cannot distinguish between real and synthetic samples.

\begin{figure}[t]
\centering
\includegraphics[scale=.35]{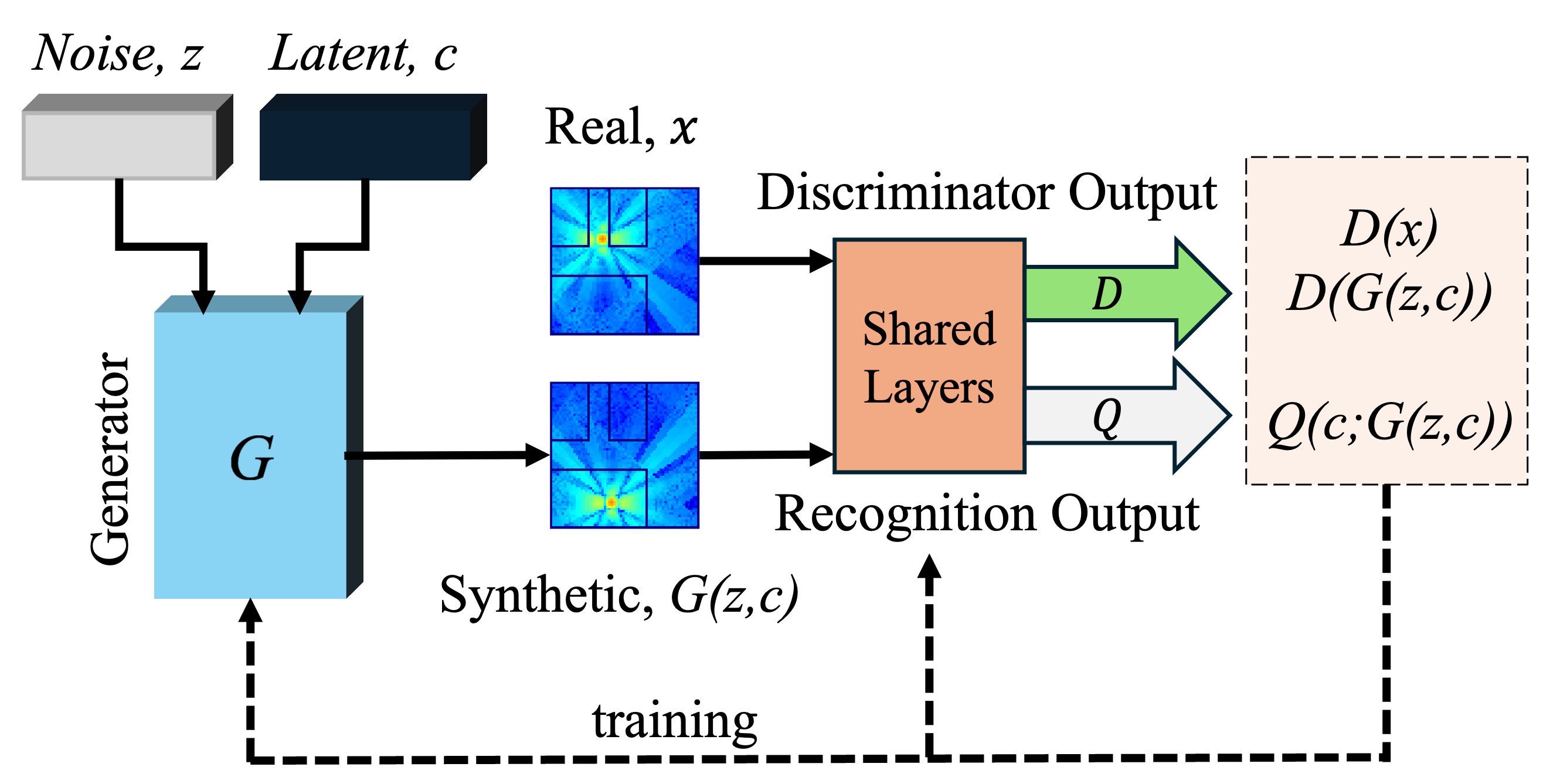}
\caption{InfoGAN architecture, comprising a generator $G$, a discriminator $D$, and a recognition network $Q$.}
\label{fig_InfoGANArch}
\end{figure}

InfoGAN~\cite{Chen2016a} extends the basic GAN framework by introducing an explicit latent variable $\cv$, which is intended to capture semantically meaningful and interpretable factors of variation in the data.
Unlike conditional GANs, $\cv$ is \emph{not} provided as labeled side information during training.
Instead, InfoGAN encourages the generator to encode information about $\cv$ into the generated sample $G(\zv,\cv)$ by maximizing the mutual information between $\cv$ and the generated data.
The resulting objective function is
\begin{equation}
\min_{G}\max_{D}\;
\Big( V(G,D) - \lambda\, I(\cv;\,G(\zv,\cv)) \Big),
\label{eq:minimaxInfoGAN_prelim}
\end{equation}
where $\lambda>0$ controls the strength of the mutual-information regularization term.
The mutual information between the latent variable $\cv$ and $G(\zv,\cv)$ is defined as
\begin{align}
I(\cv;G(\zv,\cv))
&= H(\cv) - H(\cv|G(\zv,\cv)) \nonumber\\
&= \mathbb{E}_{x\sim P_g}
\Big[\mathbb{E}_{\cv'\sim P(\cv|x)} \log P(\cv'|x)\Big] + H(\cv),
\label{eq:mi_start}
\end{align}
where $H(\cdot)$ denotes entropy, and $P(\cv|x)$ is the posterior distribution of the latent variable given a generated sample.
Here, $\cv'$ represents a latent variable drawn from this posterior, i.e., the latent value that most likely generated the sample $x$.
Direct evaluation of $I(\cv;G(\zv,\cv))$ is intractable because the true posterior $P(\cv|x)$ is unknown.
To address this, InfoGAN introduces an auxiliary distribution $Q(\cv|x)$, parameterized by a recognition network $Q$, to approximate the true posterior.
By adding and subtracting $\log Q(\cv'|x)$ and invoking the non-negativity of the Kullback--Leibler (KL) divergence, a variational lower bound on the mutual information can be derived:
\begin{align}
I(\cv;G(\zv,\cv))
&\ge
\mathbb{E}_{x\sim P_g}
\Big[\mathbb{E}_{\cv'\sim P(\cv|x)} \log Q(\cv'|x)\Big]
+ H(\cv).
\label{eq:mi_lb}
\end{align}

Using the reparameterization identity (Lemma~5.1 in~\cite{Chen2016a}),
\begin{equation}
\mathbb{E}_{x\sim P_g}
\big[\mathbb{E}_{\cv'\sim P(\cv|x)} f(\cv',x)\big]
=
\mathbb{E}_{\cv\sim P_{\cv},\,\zv\sim P_{\zv}}
\big[f(\cv,G(\zv,\cv))\big],
\label{eq:lemma51}
\end{equation}
this lower bound can be estimated directly from samples drawn from the latent prior and the noise distribution.
The resulting mutual-information regularization term is
\begin{align}
L_I(G,Q)
&=
\mathbb{E}_{\cv\sim P_{\cv},\,x\sim P_g}
\big[\log Q(\cv|x)\big] \nonumber\\
&+ H(\cv)
\le I(\cv;G(\zv,\cv)).
\label{eq:LI}
\end{align}
Since $H(\cv)$ is constant for a fixed prior $P_{\cv}$, it is omitted during optimization.
The term $L_I(G,Q)$ can be efficiently maximized with respect to $Q$ and $G$ using standard backpropagation, without modifying the overall GAN training procedure.
As $Q(\cv|x)$ approaches the true posterior $P(\cv|x)$, the variational bound becomes tight and the mutual information between $\cv$ and $G(\zv,\cv)$ is fully preserved.

In this work, we adopt the InfoGAN principle as the foundation of SAIL to learn interpretable latent factors from RF maps in a fully unsupervised manner.
By structuring the latent variable $\cv$ to capture spatial factors associated with the Tx location and angular factors associated with the Tx boresight direction, SAIL enables controlled RF map synthesis at inference time through direct manipulation of the learned latent variables.
\section{\pname{}: Architecture and Design}
\label{sec:SAIL} 

\subsection{\pname{} Architecture}
\label{sub_\pname{}}

\begin{figure*}[t]
\centering
\includegraphics[scale=.45]{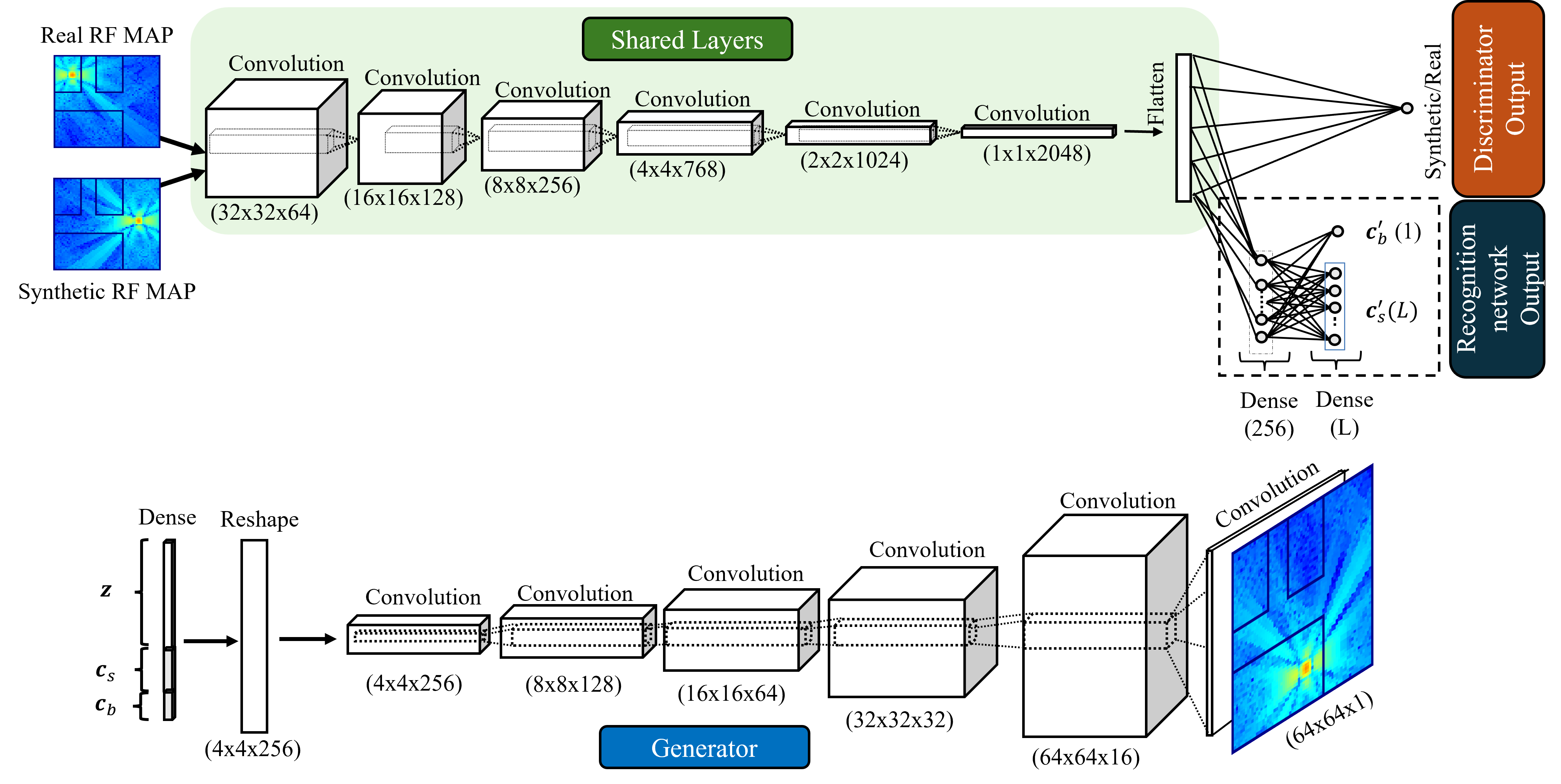}
\caption{Generator ($G$), discriminator ($D$), and recognition network ($Q$) in \pname{}. The notation $(m\times m\times n)$ below each block denotes the tensor shape in the format (height, width, channels).}
\label{f:GEN_Disc_Model}
\end{figure*}

The \pname{} architecture extends the InfoGAN framework introduced in Section~\ref{sec:Preliminaries} by explicitly structuring the latent space to capture physically meaningful spatial and angular factors in RF maps.
It consists of three neural components: a generator $G$, a discriminator (critic) $D$, and a recognition network $Q$, whose roles follow the standard InfoGAN formulation.

The generator $G$ takes as input a concatenation of a noise vector $\zv$ and a structured latent variable $\cv$, and produces a synthetic RF map $\tilde{x}=G(\zv,\cv)$.
The latent variable is defined as $\cv \overset{\scriptstyle{\text{def}}}{=} \{\cv_s, \cv_b\}$,
where $\cv_s$ is a categorical latent variable and $\cv_b$ is a continuous latent variable.
The categorical component $\cv_s$ is represented as a one-hot vector of length $L$,
$\cv_s \in \{\cv_s^1,\cv_s^2,\ldots,\cv_s^L\}$,
with each category corresponding to a distinct Tx region within the floor plan.
These regions are discovered directly from the RF maps in an unsupervised manner and serve as a coarse representation of Tx location.
During training, $\cv_s$ is sampled from a categorical prior, which is taken to be uniform over the $L$ regions.
The continuous component $\cv_b$ captures angular variations associated with the Tx beam boresight direction.
It is sampled from a uniform prior, $\cv_b \sim U(-1,1)$, which supports smooth latent traversals and enables continuous control of the synthesized RF maps with respect to the Tx boresight direction.
The noise vector $\zv \sim \mathcal{N}(\mathbf{0},\mathbf{I})$ accounts for residual stochastic variability in the RF maps that is not explained by $\cv$.

The discriminator $D$ is trained to distinguish between real RF maps drawn from $P_{\text{data}}$ and synthetic RF maps drawn from the generator distribution $P_g$.
The recognition network $Q$ shares the convolutional feature-extraction backbone with $D$ and branches at the final fully connected layers.
Given an input RF map, $Q$ outputs an estimate of the latent variables, denoted by $\hat{\cv}=\{\hat{\cv}_s,\hat{\cv}_b\}$.
Specifically, $\hat{\cv}_s$ is produced via a softmax output, while $\hat{\cv}_b$ is produced via a linear output layer.

By maximizing the mutual information between the structured latent variable $\cv$ and the generated RF map, \pname{} encourages the generator to embed spatial (Tx location) and angular (Tx boresight direction) factors directly into the synthesized RF maps.
As a result, the learned latent variables are both interpretable and directly controllable at inference time, enabling targeted RF map generation for specific spatial and angular configurations.

\subsection{\pname{} Objective Function}
\label{sub_Train}

To train \pname{}, we adopt the WGAN-GP formulation~\cite{gulrajani2017improved} to approximate the Wasserstein-1 distance between the real RF-map distribution $P_{\text{data}}$ and the generator distribution $P_g$, while enforcing the 1-Lipschitz constraint on the critic $D(\cdot)$.
Accordingly, the critic is optimized by maximizing
\begin{equation}
\begin{aligned}
V^*(G,D) \overset{\scriptstyle{\text{def}}}{=} \;&
\mathbb{E}_{\tilde{x} \sim P_g}\!\left[D(\tilde{x})\right]
- \mathbb{E}_{x \sim P_{\text{data}}}\!\left[D(x)\right] \\
&+ \beta\, \mathbb{E}_{\hat{x} \sim P_{\hat{x}}}
\!\left[\left(\|\nabla_{\hat{x}} D(\hat{x})\|_2 - 1\right)^2\right],
\end{aligned}
\label{eq:loss_WG_GP_SAIL}
\end{equation}
where $\tilde{x}=G(\zv,\cv)$ denotes a synthetic RF map and
$\hat{x}=\epsilon x + (1-\epsilon)\tilde{x}$ with $\epsilon \sim U(0,1)$ denotes an interpolated sample drawn from $P_{\hat{x}}$.
The gradient-penalty term enforces the 1-Lipschitz constraint by encouraging $\|\nabla_{\hat{x}}D(\hat{x})\|_2 \approx 1$ along straight-line paths between real and generated RF maps.

To promote disentangled and interpretable latent factors, we incorporate the mutual-information regularization term from Section~\ref{sec:Preliminaries} (Eq.~\eqref{eq:minimaxInfoGAN_prelim}) via its variational lower bound $L_I(G,Q)$.
The resulting \pname{} objective is
\begin{equation}
\begin{aligned}
\min_{G}\,\max_{D}\;
&V_{\text{SAIL}}(D,G,Q)
=\;
\mathbb{E}_{\tilde{x}\sim P_g}\!\left[D(\tilde{x})\right]
- \mathbb{E}_{x\sim P_{\text{data}}}\!\left[D(x)\right] \\
&+ \beta\,\mathbb{E}_{\hat{x}\sim P_{\hat{x}}}
\!\left[\left(\|\nabla_{\hat{x}}D(\hat{x})\|_2-1\right)^2\right]
- \lambda\,L_I(G,Q),
\end{aligned}
\label{eq:minimaxSAIL}
\end{equation}
where $L_I(G,Q)$ denotes the variational lower bound on the mutual information between the structured latent variable $\cv$ and the generated RF map.
In our implementation, we set $\beta=10$, $\lambda=2$, $L=8$, and $\dim(\zv)=5$.


\section{Ray-Tracing-Based Simulated Dataset}
\label{sec:dataset}
We generate a simulated dataset using a MATLAB-based ray-tracing engine that implements a shoot-and-bouncing-ray (SBR) method to model line-of-sight (LOS) and reflected propagation paths between a transmitter (Tx) and receivers (Rxs).
The simulated environment consists of a square indoor floor plan of size $20\,\mathrm{m} \times 20\,\mathrm{m}$, with walls modeled as brick surfaces. The relative permittivity and conductivity of the walls are set to $\epsilon_r = 3.75$ and $\sigma = 0.038$~S/m, respectively.

The Tx is equipped with an $8 \times 8$ uniform planar array (UPA) operating at a carrier frequency of 28~GHz.
The Rx antenna is assumed to be isotropic.
The Tx height is fixed at 3~m, while the Rxs are placed at a height of 1.5~m.
To capture angular variability, the Tx boresight direction is swept in the azimuth plane over $360^\circ$ with a step size of $36^\circ$, resulting in 10 distinct boresight orientations per Tx location.

\begin{figure}[t]
    \centering
    \begin{subfigure}[b]{0.45\linewidth}
        \centering
        \includegraphics[scale=1.05]{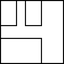}
        \caption{Indoor floor plan}
        \label{fig:sim_floorplan}
    \end{subfigure}
    \hfill
    \begin{subfigure}[b]{0.5\linewidth}
        \centering
        \includegraphics[scale=0.25]{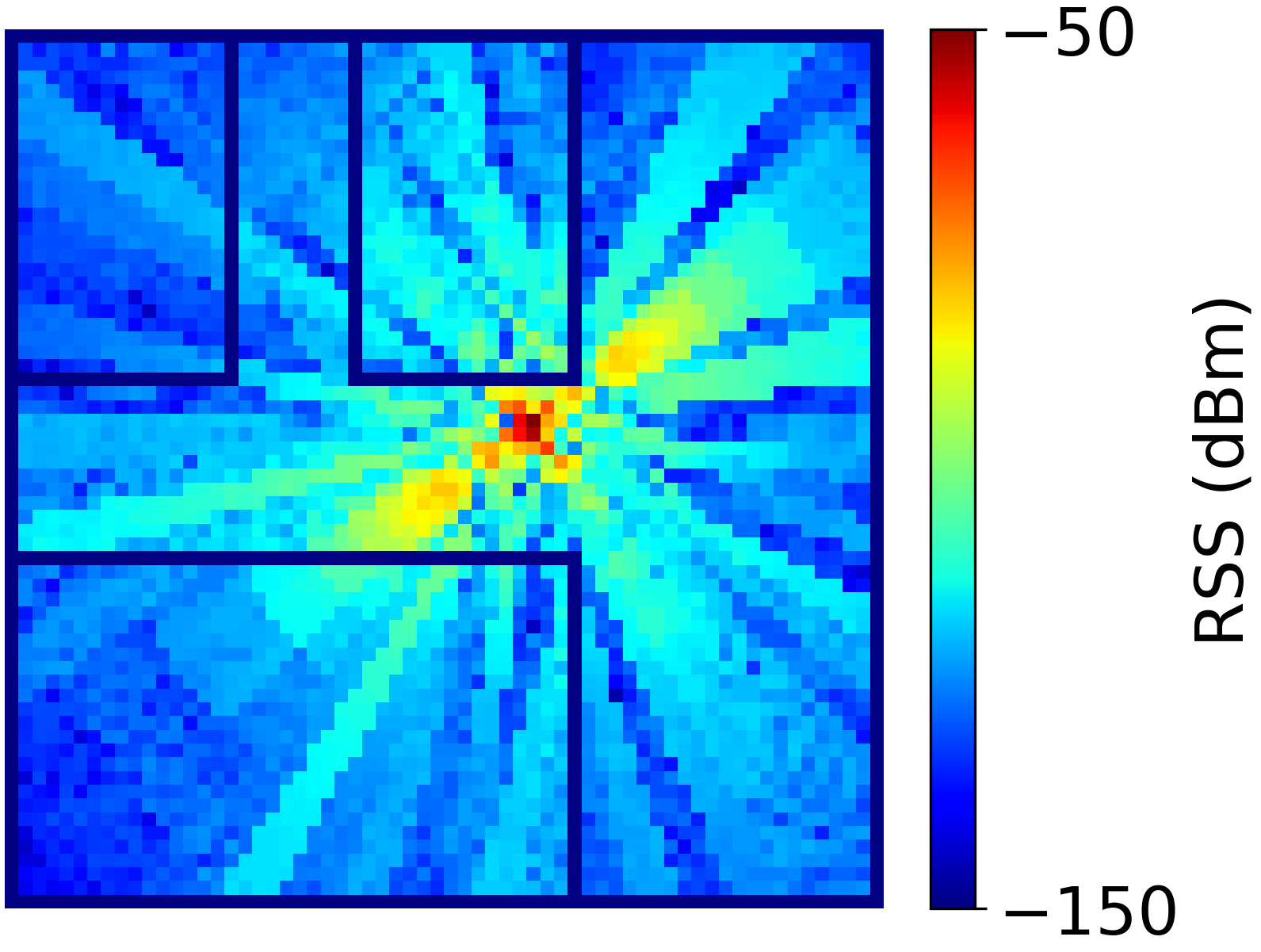}
        \caption{Example RF map}
        \label{fig:sim_rfmap}
    \end{subfigure}
    \caption{Ray-tracing-based simulated dataset: (a) the indoor floor plan of dimension $20 \times 20~\text{m}^2$ and (b) a corresponding RF map generated using $8\times8$ UPA at 28 GHz, with the Tx boresight direction at an azimuth angle of $36^\circ$.}

    \label{fig:sim_dataset_example}
\end{figure}

To construct each RF map, we evaluate the RSS over a fixed $64\times64$ spatial raster covering the floor plan, where each pixel corresponds to a receiver sampling point.
At each sampling point, RSS is computed by aggregating the received power contributions of the dominant propagation paths, including antenna gain effects and reflection losses.
We simulate $512$ distinct Tx locations distributed across the floor plan.
For each Tx location, we generate $10$ RF maps corresponding to different boresight angles, yielding a total of $5120$ RF maps.
Fig.~\ref{fig:sim_dataset_example} shows the floor plan used in the simulation and a representative RF map instance.

\section{Performance Evaluation}
\label{sec:Evaluation}

The objective of \pname{} is to learn interpretable latent factors from unlabeled RF maps and enable controllable generation by manipulating these factors at inference time. 
Accordingly, we evaluate \pname{} along three axes: (i) spatial interpretability and controllability of the categorical latent variable $\cv_s$ with respect to Tx location, (ii) angular interpretability and controllability of the continuous latent variable $\cv_b$ with respect to Tx boresight rotation, and (iii) fidelity of generated RF maps relative to ray-tracing-based simulated ground truth.

\begin{figure*}[t]
    \centering
    \includegraphics[width=0.9\linewidth]{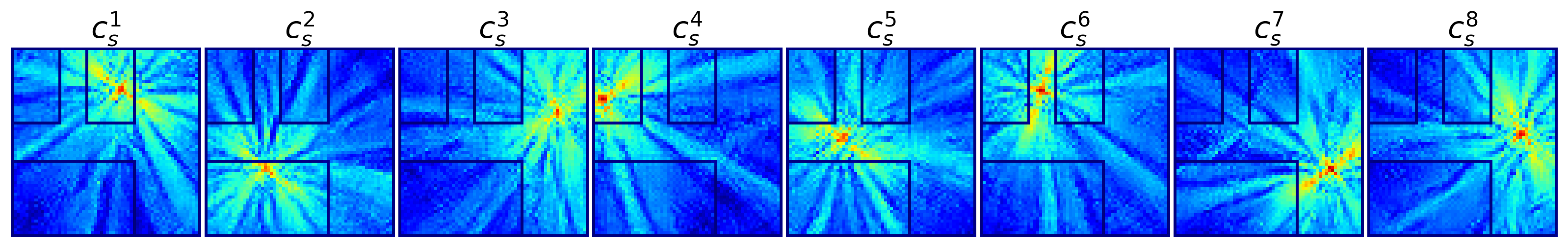}
    \vspace{-0.25cm}
    \caption{RF maps generated by \pname{} with fixed $\zv$ and $\cv_b$ while varying $\cv_s$.}
    \label{fig:cs_row}
\end{figure*}

\begin{figure}[t]
    \centering
    \vspace{-0.2cm}
    \includegraphics[scale=0.27]{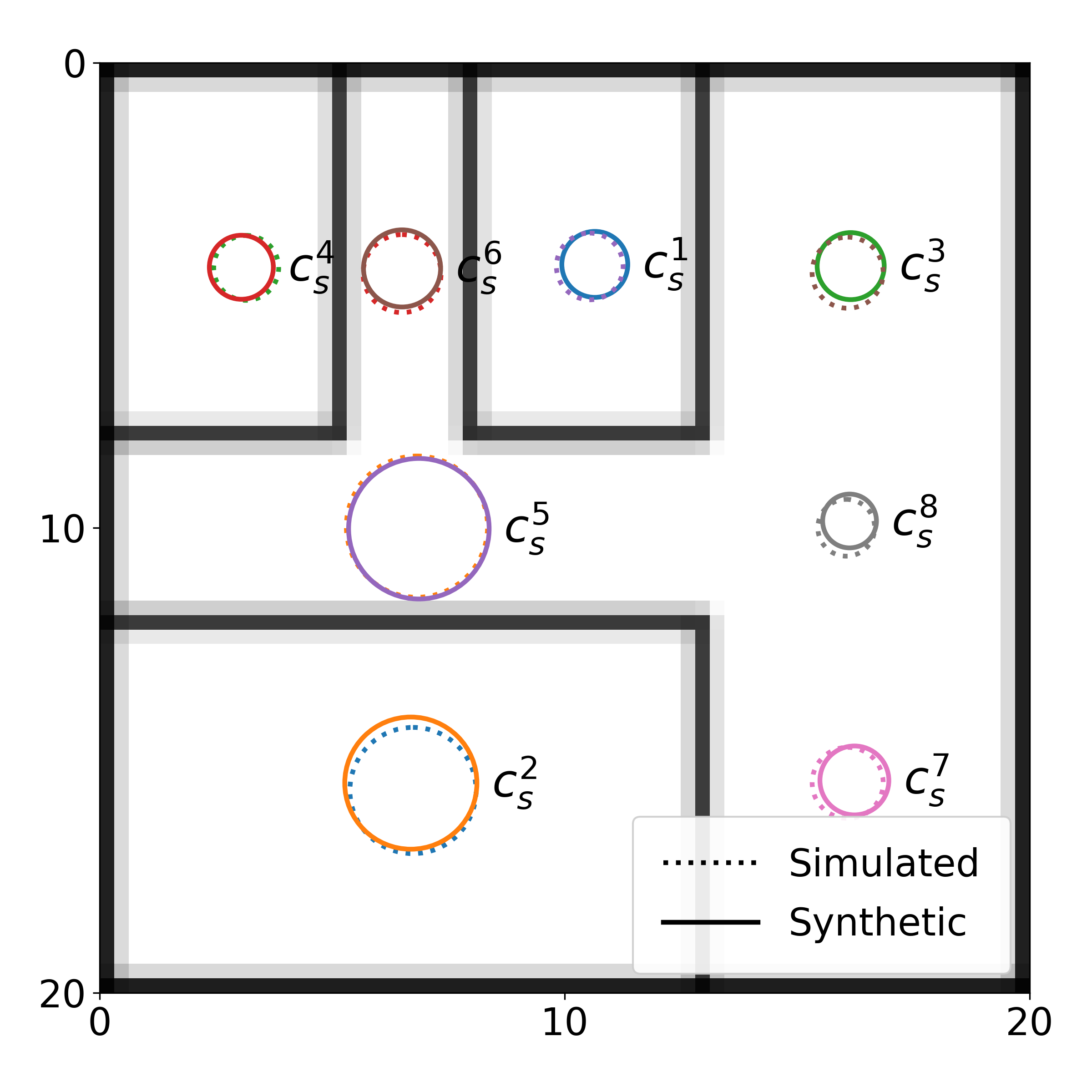}
    \vspace{-0.25cm}
    \caption{Tx locations for each region of the floor plan, shown for both synthetic (solid) and simulated (dashed) RF maps.}
    \label{fig:cs_circles}
\end{figure}

\subsection{Learning and Controlling Tx Location via $\cv_s$}
\label{subsec_spatial_control}

We evaluate whether \pname{} learns a spatially interpretable categorical latent variable $\cv_s$ that captures Tx-location-dependent propagation structure and enables controlled RF map synthesis across different regions of the floor plan.
In \pname{}, different values of $\cv_s$ are expected to correspond to distinct Tx regions learned directly from unlabeled RF maps.
By manipulating $\cv_s$ while keeping the noise vector $\zv$ and the continuous latent variable $\cv_b$ fixed, \pname{} can generate RF maps associated with specific spatial regions.
To examine this behavior, we generate synthetic RF maps by varying $\cv_s$ across its one-hot states.
Fig.~\ref{fig:cs_row} shows one representative RF map for each category.
As $\cv_s$ changes, the dominant high-power region of the RF map shifts discretely across the floor plan, suggesting that each category corresponds to a different Tx-associated spatial region.

To quantitatively assess this correspondence, we estimate an effective Tx location for each RF map.
Specifically, for a given RF map, we identify all pixels whose received power exceeds the 99th percentile of that map and compute the average of their $(x,y)$ coordinates.
This procedure yields one Tx-location estimate per RF map.
We repeat this process over multiple RF maps generated for each category of $\cv_s$, resulting in a set of Tx-location estimates for each category.

For each category, we summarize these estimates by their mean Tx location and a dispersion radius defined as
\[
r = \tfrac{1}{2}\left(\sigma_x + \sigma_y\right),
\]
where $\sigma_x$ and $\sigma_y$ denote the standard deviations of the estimated Tx-location coordinates along the horizontal and vertical axes, respectively.

Fig.~\ref{fig:cs_circles} visualizes these statistics on the floor plan.
For each category, a circle is drawn with its center at the mean Tx location and radius equal to $r$.
Solid circles correspond to statistics computed from \pname{}-generated RF maps, while dashed circles correspond to those computed from ray-tracing simulated RF maps.
Across all categories, the mean Tx locations inferred from synthetic RF maps closely match those obtained from ray-tracing data, and the corresponding dispersion radii are comparable.
This indicates that \pname{} consistently assigns distinct and well-separated spatial regions to different values of $\cv_s$, despite being trained without Tx-location labels.

Overall, these results demonstrate that $\cv_s$ captures discrete Tx-location-dependent propagation regimes and enables reliable, label-free control of Tx location in synthesized RF maps.

\subsection{Learning and Controlling Boresight Direction via $\cv_b$}
\label{subsec_angular_control}

We next evaluate whether \pname{} learns an angularly interpretable continuous latent variable $\cv_b$ that captures smooth variations in Tx boresight direction in RF maps.
To isolate angular behavior from other generative factors, we fix $\zv$ and $\cv_s$, and vary only $\cv_b$ over its support $[-1,1]$, generating RF maps along a continuous latent traversal.

For each generated (and simulated) RF map, the Tx location is first estimated using the procedure described in Section~\ref{subsec_spatial_control}.
Conditioned on this estimated Tx location $(x_0,y_0)$, we then estimate the Tx boresight direction from the spatial distribution of the strongest RSS region.
The underlying intuition is that, for directional transmissions, the high-RSS region forms an elongated structure aligned with the Tx boresight direction.

Let $\Omega = \{(x,y) : P(x,y)\ge \tau\}$ denote the set of spatial points whose RSS exceeds a high threshold $\tau$, where $P(x,y)$ is the RSS at location $(x,y)$.
In our experiments, $\tau$ is set to the $95$th percentile of the RSS values to ensure robustness while retaining sufficient spatial support.
We compute a weighted covariance matrix of the high-RSS region relative to the estimated Tx location as
\begin{equation}
\mathbf{C}
=
\sum_{(x,y)\in\Omega}
w(x,y)
\begin{bmatrix}
x-x_0 \\[2pt]
y-y_0
\end{bmatrix}
\begin{bmatrix}
x-x_0 & y-y_0
\end{bmatrix},
\label{eq:boresight_cov}
\end{equation}
where the weights are proportional to the RSS,
$w(x,y)\propto P(x,y)$, and normalized such that
$\sum_{(x,y)\in\Omega} w(x,y)=1$.
The principal eigenvector $\mathbf{v}=[v_x,v_y]^T$ of $\mathbf{C}$ identifies the dominant spatial axis of the high-RSS region and is taken as an estimate of the Tx boresight direction.
The corresponding boresight angle is computed as
\begin{equation}
\theta_b = \mathrm{atan2}(-v_y,\,v_x),
\label{eq:boresight_angle}
\end{equation}
where the negative sign accounts for the image coordinate system, in which the vertical axis increases downward.
The inherent sign ambiguity of principal component analysis is resolved by selecting the direction along $\pm\mathbf{v}$ that yields higher average RSS when sampling forward from $(x_0,y_0)$.

\begin{figure*}[t]
    \centering
    \includegraphics[width=\textwidth]{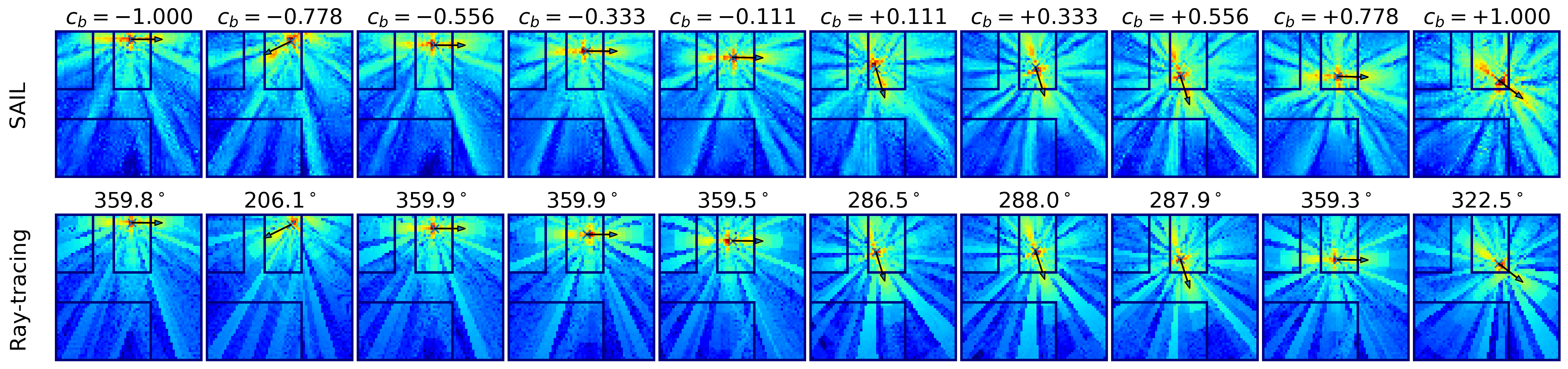}
    \caption{Boresight direction control via latent traversal of $\cv_b$ for $\cv_s^5$. 
    Top: RF maps synthesized using \pname{} with fixed $(\zv,\cv_s)$ and $\cv_b$ swept from $-1$ to $+1$. Bottom: ray-tracing simulated RF maps.
    Arrows indicate the estimated dominant direction.}

    \label{fig:cb_control}
\end{figure*}

Fig.~\ref{fig:cb_control} illustrates the resulting angular control for a representative category $\cv_s^5$.
We fix $(\zv,\cv_s)$ and vary $\cv_b$ uniformly from $-1$ to $+1$, generating the synthetic RF maps shown in the top row.
For each synthetic map, we retrieve a ray-tracing RF map whose estimated Tx location and boresight direction most closely match those of the synthetic sample.
The retrieved ray-tracing RF maps are shown in the bottom row, with titles indicating the estimated boresight angles.

As $\cv_b$ varies, the dominant lobe in the synthetic RF maps rotates smoothly, and the retrieved ray-tracing RF maps exhibit consistent boresight orientations and similar spatial RSS distributions.
These results demonstrate that \pname{} learns a continuous and interpretable angular latent variable, enabling explicit control of Tx boresight direction through $\cv_b$ without any angle supervision during training.

\subsection{RF Map Fidelity Relative to Ray Tracing}
\label{subsec_fidelity}

We evaluate whether \pname{} generates RF maps that are statistically consistent with ray-tracing RF maps.
After training, we synthesize RF maps by sampling $(\zv,\cv_s,\cv_b)$ from their respective priors.
Because \pname{} is trained without labels, we first establish a correspondence between each learned categorical mode and a physical region (room) in the simulated floor plan.
This correspondence is inferred based on proximity between the average Tx locations estimated from synthetic RF maps and those associated with ray-tracing data.
Using this mapping, we form ray-tracing--synthetic RF map pairs as follows.
For each ray-tracing RF map, we select a synthetic RF map from the corresponding category whose inferred Tx location and boresight direction most closely match the ray-tracing configuration.
This matching procedure avoids comparisons between RF maps generated under incompatible propagation geometries and enables a physically meaningful fidelity assessment.
Using the resulting matched pairs, we compute standard image-based similarity metrics between ray-tracing RF maps and their synthetic counterparts.
Averaged across all rooms, \pname{} achieves an SSIM of $0.8576$ and a PSNR of $23.33$\,dB.
These results indicate that the generated RF maps preserve the dominant spatial structures induced by path loss, blockage, and reflections, despite the absence of explicit supervision during training.

\section{Conclusions}
\label{sec:conclusions}

This paper presented \pname{}, a generative adversarial framework that learns a structured, interpretable latent space directly from unlabeled RF maps without requiring explicit annotations of Tx location or boresight direction, enabling targeted and fine-grained control at inference time through latent traversal.
Using a ray-tracing-based indoor dataset at mmWave frequencies, we demonstrated that \pname{} generates RF maps that are statistically consistent with ray-tracing simulations, achieving an average SSIM of $0.8576$ and an average PSNR of $23.33$\,dB.
We further showed that the categorical latent variable partitions the floor plan into distinct Tx-dependent regions, while the continuous latent variable captures smooth and physically meaningful variations in Tx boresight direction.
These results confirm that \pname{} discovers propagation-relevant factors and enables systematic manipulation of spatial and angular effects despite fully unsupervised training.
Future work will extend \pname{} to dynamic and multi-Tx environments, incorporate over-the-air experimental measurements, and explore integration with downstream tasks such as beam selection, beam tracking, and localization.

\bibliographystyle{IEEEtran}
\bibliography{bibfile}
\end{document}